\documentstyle[pra,aps]{revtex}
\begin{document}
\draft
\author{Li-Bin Fu$^1$, Jie Liu$^{1,2}$ , Jing Chen $^{2,3}$ and Shi-Gang Chen%
$^1$}
\address{$^1$Institute of Applied Physics and Computational Mathematics,\\
P.O. Box 8009 (26), 100088 Beijing, China\\
$^2$CCAST (World Laboratory), P.O.Box 8730, Beijing\\
$^3$Institute of Theoretical Physics, Chinese Academy of Science\\
P.O. Box 2735, Beijing 100080, China}
\title{Classical Collisional Trajectories as the Source of Strong Field Double
Ionization of Helium in the Knee Regime}
\maketitle

\begin{abstract}
\begin{center}
{\bf Abstract }
\end{center}

In this paper, a quasistatic model is extended to describe the double
ionization of Helium in intense linearly polarized field, yielding achieve
an insight to the two-electron correlation effect in the ionization
dynamics. Our numerical calculations reproduce the excessive double
ionization and the photoelectron spectra observed experimentally both
quantitatively and qualitatively. Moreover, it is shown that the classical
collisional trajectories are the main source of the double ionization in the
knee regime and responsible for the unusual angular distribution of the
photoelectrons.
\end{abstract}


Recently the excessive double ionization observed in Helium experiments by
Fittinghoff et al. \cite{cet3}, Walker et al. \cite{cet4}, and Sheehy et al.
\cite{cet5} draws much attention to the multiple-electron dynamics in the
laser-atom interaction. In these experiments the single ionization yields of
He in a linearly polarized field is accurately predicted by the single
active electron (SAE) approximation \cite{cet4}, well described by the
Ammosov-Delone-Krainov (ADK) tunneling theory \cite{ndi5}. However, the case
of double ionization is more complicated. In the regime of very high
intensities ($I>10^{16}$W/cm$^2$) where strong double ionization occurs, the
double ionization keeps in good agreement with the sequential SAE models as
that in the lower intensities regime($I<10^{14}$W/cm$^2$). The double
ionization yield deviates seriously from the sequential SAE model and shows
a great enhancement in a ''knee'' regime $[(0.8$---$3.0)\times 10^{15}$W/cm$%
^2],$ where the He$^{2+}/$He$^{+}$ yields ratio is close to a constant:
0.002. This surprising large yields of the double ionization obviously
indicates that the sequential ionization is no longer the dominating process
in this regime and the electron-electron correlation has to be taken into
account.

Both the ``shake-off'' model and the ``recollision'' model are suggested to
describe the electron's correlation \cite{cet3,cet5,cfi2,m5}. However, none
of the two nonsequential ionization (NSI) mechanisms can completely explain
the experimental observations. For the ``shake-off'' model, it can not give
the reason for the decrease in the double ionization yields as the
polarization of the laser field departs from linear \cite{ee28,ee34,ee35}.
In the ``recollision'' model, the returning electrons are known to have a
maximum classical kinetic energy of $\sim 3.2U_p$ $(U_p=e^2F^2/4m_e\omega
^2) $, so one can determine a minimum intensity required for the
rescattering electron to have enough energy to excite the inner electron.
But the double ionization yields observed in experiments have no such an
intensity threshold. In fact, the double ionization process is rather
complicated and subtle, both of the two NSI processes and the sequential
ionization contribute to the double ionization yields and may dominate in
the different regimes.

The experiments on the double ionization of Helium are mainly confined in
the tunneling regime, i.e. the ratio between the tunneling time of the outer
electron and the inverse optical frequency (Keldysh parameter) is less than
1. In this regime, the quasistatic model \cite{cfi2} provides a perfect
description for the Hydrogen-like atoms in the intense fields and
successfully explain the most nonlinear phenomena observed experimentally%
\cite{cfi2,hbbown,chenjing}. Inspired by this success, in this paper we
extend to develop a 3D quasistatic model (a two step process) to investigate
the mechanism of the double ionization of Helium by tracing the classical
trajectories of the two correlated electrons. We attribute the double
ionization to the classical collisional trajectories: Distinct trajectory
configurations corresponding to the ''shake-off'' and ``recollision''
mechanism contribute to the nonsequential double ionization of helium. Our
numerical simulations successfully reproduce the excessive double ionization
and the photoelectron spectra observed experimentally in the knee regime .
The intuitionistic pictures of the double ionization will be provided by
this model.

As a beginning, we present the improved two-step quasistatic model adopted
in our calculations. The first step that the outer electron tunnels free, is
treated by the tunneling ionization theory generalized by Delone et al. \cite
{chj21}. In the second step, the evolution of the two electrons after the
first electron tunneled and the electron-electron interaction are described
by the classical equations (in atomic unit):
\begin{equation}
\frac{d^2{\bf r}_1}{dt^2}=-\frac{2{\bf r}_1}{r_1^3}+\frac{{\bf r}_1-{\bf r}_2%
}{|{\bf r}_1-{\bf r}_2|^3}-{\bf F}(t),  \label{eq1}
\end{equation}
\begin{equation}
\frac{d^2{\bf r}_2}{dt^2}=-\frac{2{\bf r}_2}{r_2^3}-\frac{{\bf r}_1-{\bf r}_2%
}{|{\bf r}_1-{\bf r}_2|^3}-{\bf F}(t),  \label{eq2}
\end{equation}
where ${\bf F}(t)$ is the laser field.

In our model, the initial state of the inner electron of the Helium is
described by a microcanonical distribution which is widely used in the
classical-trajectory Monte Carlo (CTMC) methods established and developed by
\cite{rbp421u,bdh8}. The CTMC method has been successfully used in studying
the interaction of atoms with strong laser fields by numerous authors \cite
{rbp421,bdh}, which provides a statistical distribution of all the
parameters defining the initial conditions of a trajectory of the electrons
in the ground state of a hydrogen-like atom. Then, the initial distribution
of the inner electron is
\begin{equation}
\rho ({\bf r}_2,{\bf p}_2)=\frac{\delta (E_2-H_0(r_2,p_2))}K,  \label{dis}
\end{equation}
where $H_0(r_2,p_2)=p_2^2/2m_e-Ze^2/r_2$; $K$ is the normalization constant;
$E_2=-2$ $a.u.$ is the eigenenergy of the inner electron. Integrating the
above equation, one obtains the momentum distribution
\begin{equation}
\rho ({\bf p}_2)=\frac{8p_c^5}{\pi ^2(p_2^2+p_c^2)^4},  \label{dis1}
\end{equation}
in which $p_c^2=2m_eU,$ $U$ is the negative energy of the inner electron.

The spherically symmetric ground-state He$^{+}$ is represented by the above
microcanonical distribution. This state is specified by the binding energy
of the electron in the target atom and five additional parameters randomly
distributed in the following ranges: $-\pi \leq \phi \leq \pi ,$ $-1\leq
\cos \theta \leq 1,$ $-\pi \leq \eta <\pi ,$ $0\leq \epsilon ^2\leq 1$ and $%
0\leq \chi _n\leq 2\pi $ (Ref. \cite{bdh8}). Here, $\epsilon $ is the
eccentricity of the orbit, $\chi _n$ is a parameter of the orbit
proportional to time, and $\phi ,$ $\theta $ and $\eta $ are Euler angles. A
random distribution of these parameters corresponds to equal probability of
the inner electron having any phase in its periodic motion. Here, $10^4$
initial points are chosen and their momentum distribution is compared with
Eq. (\ref{dis1}). Figure 1 shows that they are agreeable.

The initial condition of the tunneled electron, under the SAE approximation
of He$^{+}$, is determined by a equation including the effective potential
given in Ref. \cite{chj22} and a generalized tunneling formula developed by
Delone et al. \cite{chj21}. In parabolic coordinates, the Schr\"odinger
equation for a hydrogen-like atom in a uniform field $\epsilon $ is written
(in atomic unit),
\begin{equation}
\frac{d^2\phi }{d\eta ^2}+(\frac{I_{p1}}2+\frac 1{2\eta }+\frac 1{4\eta ^2}+%
\frac 14\epsilon \eta )\phi =0,  \label{sch}
\end{equation}
in which $I_{p1}=-0.9$ $a.u.$ is the negative ionization potential of the
outer electron.

The above equation has the form of the one-dimensional Schr\"odinger
equation with the potential $U(\eta )=-1/4\eta -1/8\eta ^2-\epsilon \eta /8$
and the energy $K=\frac{I_{p1}}4.$ The turning point, where an electron born
at time $t_0$, is determined by $U(\eta )=K$. In the quasistatic
approximation, the above field parameter $\epsilon $ relates to the laser
field amplitude $F(t)$ by $\epsilon =F(t_0)$. One must point out, as $%
\epsilon >F_{th},$ the turning point will be complex, which determines the
threshold value of the field $F_{th}=0.338$ $a.u.$

The evolution of the outer electron is traced by launching a set of
trajectories with different initial parameters $t_0$ and $v_{1x0}$, where $%
v_{1x0}$ is the initial velocity perpendicular to the polarization of the
electric field. The initial position of the electron born at time $t_0$ is
given by $x_{10}=y_{10}=0$ , $z_{10}=-\eta _0/2$ form the Eq. (\ref{sch}).
The initial velocity is set to be $v_{1y0}=v_{1z0}=0,$ $v_{1x0}=v_{10}$.
Thus, the weight of each trajectory is evaluated by \cite{chj21}
\begin{equation}
w(t_0,v_{10})=w(0)w(1),  \label{wei1}
\end{equation}
\begin{equation}
w(1)=\frac{\sqrt{2I_{p1}}v_{10}}{\epsilon \pi }\exp (-\sqrt{2I_{p1}}%
v_{10}^2/\epsilon ),  \label{wei2}
\end{equation}
and where $w(0)$ is the tunneling rate in the quasistatic approximation \cite
{cfi11}.

Before we go further, we would like to compare our model with a similar
model \cite{cfiown} describing the double ionization of helium. First, in
our model the initial condition of the inner electron is given by the
classical trajectory Monte Carlo method (CTMC); Second, the Coulomb
interaction is described by the real Coulomb potential. These improvements
are essential. In the model given in Ref.\cite{cfiown}, the inner electron
is assumed to be rest at the center. This initial condition confines the
motion of both electrons in the same plane defined by the polarization axis
and the direction of the initial transverse momentum., i.e., in fact, the
calculations in their paper is a 2D system, which may increase the
probability of the collisions between the two electrons. On the other hand,
the soften Coulomb potential approximation adopted in Ref.\cite{cfiown}
makes the inner electrons more easily to be excited and cause an
overestimation of the double ionization rate. Our model has been employed to
understand the momentum distribution of the recoil ions and shown a good
agreement with the experimental records\cite{newnew}.

In our calculation, the Eqs. (\ref{eq1}) and (\ref{eq2}) are solved in a
time interval between $t_0$ and $13T$ by employing the standard Runge-Kuta
algorithm$.$ After ten optical cycles the electric field is switched off
using a $cos^2$ envelope during three cycles, and during the last two
optical cycles the electrons is free from the electric field. So, the
electric field can be expressed as
\begin{equation}
{\bf F}(t)=a(t)F\cos (\omega t){\bf e}_z,
\end{equation}
where $F$ and $\omega $ are the amplitude and frequency of the field
respectively and the envelope function $a(t)$ is defined by
\begin{equation}
a(t)=\left\{
\begin{array}{cc}
1 & t\leq 10T \\
\cos ^2\frac{(t-10T)\pi }{6T} & 10<t\leq 13T \\
0 & t>13T
\end{array}
\right. .
\end{equation}
The wavelength is $\lambda =780$ $nm$, which is so chosen to match the
experiment \cite{cet4}, and the intensities ranging from $I=10^{14}$ $W/cm^2$
to the threshold value $I=4\times 10^{15}$ $W/cm^2$ .

In our computations, $10^5$ or more initial points are randomly distributed
in the parameter plane $-\pi /2<\omega t_0<\pi /2,$ $v_{1x0}>0$ for the
outer electron and in the microcanonical distribution for the inner
electron. The probability for double ionization and the angular distribution
can be obtained by making statistics on an ensemble of classical
trajectories weighed by the (\ref{wei1}). The results have been tested for
numerical convergence by increasing the number of trajectories.

In our treatment, the behavior of the classical trajectories play an
important role and determine the ionization dynamics of the electrons. There
are four kinds of typical trajectories. Fig. 2(a) shows a simple behavior:
After tunneled out, the outer electron will be driven mainly by the field
and directly run away. It collides neither with core nor with the inner
electron. Fig. 2(b) gives a more complicated picture in which multiple
returns and long-time trapping is experienced by the outer electron: The
outer electron first tunneled out, and then oscillate in the combined laser
and Coulomb fields. After several optical periods, it collides with the core
and then absorb enough energy to escape. In the above two cases, no double
ionization occurs since the collision between the two electrons is slight.
Fig. 2(c) and 2(d) give the typical pictures of the double ionization
process. In Fig. 2(c), the outer electron is born at the regime close to the
peak of the electric field, then it oscillates in the combined laser and
Coulomb fields. After several optical periods, it returns back to the
neighborhood of the core and collides strongly with the inner electron. This
collision provide enough energy for the inner bounded electron to get free.
Fig. 2(d) shows that after the outer electron is tunneled, the laser field
will reverse its direction within less than a quarter of the optical
periods, so that this electron will be driven back when the laser field
reverses its direction and collides with the inner electron near the core
and make it ionized. As we will show later, the resulting energy spectra and
the angular distribution of the photoelectrons for the two processes are
quite distinct.

To match the experiments, Figure 3 shows the double ionization yields of
helium calculated by making use of our model at 13 different intensities in
the range $4\times 10^{14}-4\times 10^{15}$ $W/cm^2.$ The dashed line is the
single-ionization yields of He predicted by the ADK tunneling rate \cite
{ndi5}, and the solid line is the ADK tunneling rate for He$^{+}$. For peak
intensities below $3\times 10^{15}$ $W/cm^2$ , one sees that the double
ionization rate obtained from our numerical simulations is larger than the
ADK tunneling rate, but for the intensities above $3\times 10^{15}$ $W/cm^2$%
, the ADK tunneling rate increases rapidly and becomes larger than the
ionization rate given by our model. This figure reads that our calculation
is able to reproduce, qualitatively at least, the excessive double
ionization observed in helium experiments \cite{cet4}. The inset in Figure 3
shows the double ionization rate calculated by our model normalized to the
ADK tunneling rate of He versus the intensity. Our result is in good
agreement with the data in the knee regime observed in experiments \cite
{cet4}: He$^{2+}/$He$^{+}$ ratio in the knee regime is nearly around $0.002$%
. At lower intensities $(I<0.5\times 10^{15}/cm^2),$ the deviation between
our calculation and the experimental records becomes serious. In
conclusions, our model provides a suitable description for the double
ionization in the knee regime, where as shown above the classical
collisional trajectories (Fig.2c,d) are believed to be the main source of
the double ionization. Above this regime, the tunneling ionization of the
inner electron will become a dominating process and the ADK description is
available. Below this regime, the ionization mechanism of the outer electron
transits from tunneling regime to the multiphoton regime and the tunneling
description is no longer available.

Figure 4 shows the relations between the ionization rate and the phase of
the laser field when the outer electron tunneled. One finds that the most
double ionization yields come from the region $(-0.2<\omega t_0<0.4)$ close
to the peak of the electric field. There is a tail for the regime $\omega
t_0>0.4$ and a 'cut off' for the $\omega t_0<-0.2$. We know that when the
outer electron tunneled out near the peak of the laser field, its canonical
momentum is almost zero. Hence, the outer electron tends to oscillate in the
combined laser and Coulomb fields for several optical periods, and then
return back to the neighborhood of the core to collide with the inner
electron. In this case, the typical trajectory of the double ionization
process corresponds to Fig. 2(c). For phase $\omega t_0<-0.2$, the tunneled
electrons have a nonzero canonical momentum directing outwards from the
core. Consequently, it will be driven by the laser field and run away
directly from the core. That is, in this process, the outer electron has no
chance to return to the core and no double ionization occurs in the region.
For phase $\omega t_0>0.4$, the outer tunneled electron has a nonzero
canonical momentum towards the core , and soon after it tunneled out the
laser field also reverses its own direction to the same direction. So the
electron will be driven back to the core by the external field and collide
with the inner electron. The Fig. 2(d) shows the typical trajectory for this
case. In this region the tunnel ionization of the outer electron is not
efficient, and the double ionization rate is low. Comparing the two typical
processes of the double ionization, one can find some intrinsic difference.
In the Fig. 2(c) the outer electron was firstly ionized out, then driven by
the field to collide with the inner electron and cause the double
ionization, which shows a typical picture of the `recollision' process. In
the Fig.2(d) the inner electron was ionized during the process when the
outer electron was driven away from core by the external field, both
electrons ionize simultaneously which possesses the properties of the
`shake-off ' mechanism. The difference of the two processes manifests
clearly in the energy evolution of the two electrons. As shown in Fig. 5(a),
the outer electron is ionized free with a positive energy, then it comes
back to collide with the inner electron. This collision causes an sudden
increment on the energy of the inner electron which becomes free soon.
Because the collisions between the two electrons is almost instantaneous so
that the energy is conserved approximately when collision happens. In the
case of Fig. 5(b),during the escape process of the outer electron it collides
with the inner one. Consequently, both electrons are ionized free almost
simultaneously. From our calculation we know that both processes contribute
to the double ionization in the knee region, but the main contribution comes
form the `recollision' process which gives more than $80\%$ of the double
ionization yields.

From our calculations, we can also obtain the photoelectron spectra (PES)
and the photoelectron angular distribution (PAD). Figure 6 shows the total
photoelectron energy distribution at $1\times 10^{15}$ $W/cm^2$ and $%
1.6\times 10^{15}$ $W/cm^2$ (both of them are in the knee regime) calculated
from our model. On can see that, in absolute units, an increasing laser
intensity results in the increase of higher energy photoelectrons. But if
one scales the energy units by the pondermotive energy $U_p=e^2F^2/4m_e%
\omega ^2,$ of electron, one will find that the PES for both intensities
will show similar shape: The spectrum exhibits a sharply decreasing slope
(region {\bf I}, $0$--$2U_p$) followed an extended plateau up to $8U_p$ or
more (region {\bf II)}. This spectrum structure is much close to
experimental observations in this regime\cite{cet4}.

We know that in our model the electron is initiated in the phase interval $%
[-\pi /2,\pi /2],$ so the total photoelectron angular distribution (PAD) has
to consider the contribution of electrons originated in $[\pi /2,3\pi /2]$
which is the mirror image with respect to $90^0$. Figure 7(a) is the total
angular distribution of ionized electrons. One finds that except for a main
concentration in the field direction, there exists a long tail decreasing
with a power-law dependence $1/(\sin ^r(\theta /2))$ which is different from
the angular distribution for the pure tunneling nature where it decreases
exponentially. This structure is due to the scattering with the core and the
interaction between two electrons during the rescattering processions.
Figure 7(b) shows the angular distribution of photoelectrons in the energy
region {\bf II}. The most striking feature of the plots is the existence of
a slight slope up to $40^0$ followed by a tail up to $90^0$. If one compares
this result with the angular distribution of the transition region in
rescattering processes described in Ref. \cite{hbbown}, where there is no
photoelectrons emitted at angles much larger than $40^0,$ it is not
difficult to find that the tail structure is due to the electron-electron
interaction. Because the velocity direction of the inner electron is random
when the collision happens, the instantaneous strong interactions can give
rise to large emission angle of the photoelectron. This can be verified in
Figure 7(c), which shows the angular distributions of the inner electron and
outer electron in double ionization respectively. This case corresponds to
strong electron-electron interaction and the angular distributions is almost
flat and the decreases is slowly when the emission angle increases.

In fact, the final energy and angular distribution of the photoelectrons are
greatly determined by the scattering processes. The process shown in Fig.
2(a) only provides a relatively low energy for the photoelectron (region
{\bf I)} and gives small emission angle (the field direction). However, for
the process shown in Fig 2(b), the classical trajectories have complex
behavior, and the energy exchange in this process is also complicated. The
multiple returns and long-time trapping can produce high energy electron in
this process. So, this region contributes to the high energy part of PES{\bf %
.} It is also found that the structures of PES and PAD are similar to some
extent to the rescattering model of hydrogen \cite{hbbown}. This fact
indicates that the main structures of PES and PAD come from the rescattering
process of the outer electron with the ion He$^{+}$ .

In conclusions, a quasistatic two step model is used to investigate the
double ionization of helium in intense linearly polarized field. Our
calculations reproduce the excessive double ionization and the photoelectron
spectra observed in experiments. We argue that the classical collisional
trajectories are the main source of the double ionization in the knee regime
and responsible for the unusual angular distribution of the photoelectrons.
Two distinguished typical collisional trajectories correspond to the
`recollision' process and the `shake-off' process respectively. Both of the
two processes have contribution to the double ionization, but the
`recollision' gives the main contribution and leads to more than $80\%$ of
the double ionization yields. Our calculations based on the classical
trajectories provide an intuitionistic picture of the double ionization of
helium, and are helpful in understanding the complicated behavior of
multi-electrons atoms in intense laser fields, in the way of quantum
approach and the future experiments observations.

We acknowledge very helpful discussions with Dr. T.W. Cheng. This work was
supported by the Important Fundamental Researching Project of China.

\section{Figures caption:}

Fig. 1. Momentum distribution of the initial state of the inner electron.
the full circles denote the results of $10^4$ Monte Carlo points, the solid
line is the distribution given by the formula (\ref{dis1}).

Fig. 2. Four typical trajectories in our calculation. (a) The final energy
is $E_1=4.694$ $a.u.$and $E_2=-2.01$ $a.u.$, and the final angle of the
outer electron $\theta =2.26${\bf ; (}b{\bf ) }The final energy is $%
E_1=8.683 $ $a.u.$and $E_2=-1.153$ $a.u.$, and the final angle of the outer
electron $\theta =5.67$. (c) The typical trajectories of electrons in double
ionization corresponds to $\omega t_0$ in the phase interval $(-0.2,0.4)$.
The initial conditions are $\omega t_0=-0.087,$ the weight of the trajectory
$0.168;$ the final energy is $E_1=3.407$ $a.u.$, $E_2=3.278$ $a.u.,$ and the
final angle of two electrons: $\theta _1=25.15,$ $\theta _2=30.86$. (d) The
typical trajectories of electrons in double ionization corresponds to the
phase interval $(\omega t_0>0.4).$ Here $\omega t_0=0.924,$ the weight of
the trajectory $0.014;$ the final energy is $E_1=2.663$ $a.u.$, $E_2=0.237$ $%
a.u.$, and the final angle of two electrons: $\theta _1=19.2,$ $\theta
_2=129.0$.

Fig. 3. Numerically calculated the double ionization yields of He in our
model. The dashed and dotted lines correspond to the single ionization
yields of He and He$^{+}$ predicted by ADK tunneling ionization
respectively; the full circles correspond to the results of our calculation.
Inset: Intensity dependence of He$^{2+}$/He$^{+}$ ratio given by our model.
The solid line is gotten from the experiment \cite{cet4}.

Fig. 4. The double ionization of He versus the phase of the laser field at
the moment when the outer electron tunneled.

Fig. 5. The energies evolution of the two electrons during the double
ionization process. (a) corresponding to the case of Fig. 2(c), and (b)
corresponding to the case of Fig. 2(d). The solid line represents the outer
electron and the dashed line represents the inner electron.

Fig. 6. Photoelectron energy spectra calculated from our model.

Fig. 7. Photoelectron angular distribution at $2\times 10^{15}$ W/cm$^2.$
(a) shows the total distributions of photoelectrons; (b) is the PAD for
energy region {\bf II}. (c) PAD of the inner electron and outer electron in
double ionization.


\begin{references}
\bibitem{cet3}  D.N. Fittinghoff, P.R. Bolton, B. Chang, and K.C. Kulander,
Phys. Rev. Lett. {\bf 69}, 2642 (1992).

\bibitem{cet4}  B. Walker, B. Sheehy, L.F. DiMauro, P. Agostini, K.J.
Schafer, and K.C. Kulander, Phys. Rev. Lett. {\bf 73}, 1227 (1994)

\bibitem{cet5}  B. Sheehy, R. Lafon, M. Widmer, B. Walker, L.F. DiMauro,
P.A. Agostini, and K.C. Kulander, Phys. Rev. A {\bf 58}, 3942 (1998)

\bibitem{ndi5}  M.V. Ammosov, N.B. Delone, and V.P. Krainov, Sov. Phys. JETP
{\bf 64}, 1191 (1986)

\bibitem{cfi2}  P. Corkum, Phys. Rev. Lett. {\bf 71}, 1994 (1993)

\bibitem{m5}  K.C. Kulander, J. Cooper, and K.J. Schafer, Phys. Rev. A {\bf %
51}, 561 (1995)

\bibitem{ee28}  B. Walker, E. Mevel, B. Yang, P. Berger, J.P. Chambaret, A.
Antonetti, L.F. DiMauro, and P.A. Agostini, Phys. Rev. A {\bf 48} R894 (1993)

\bibitem{ee34}  D.L. Fittinghoff, P.B. Bolton, B. Chang, and K.C. Kuander,
Phys. Rev. A {\bf 49}, 2174 (1994)

\bibitem{ee35}  K. Kondo, A. Sagiska, T. Tamida, Y. Nabekawa, and S.
Watanabe, Phys. Rev. A {\bf 48}, R2531 (1993)

\bibitem{hbbown}  B. Hu, J. Liu and S.G. Chen, Phys. Lett. A {\bf 236}, 533
(1997)

\bibitem{chenjing}  J. Chen, J, Liu and S.G. Chen, Phys. Rev. A {\bf 61},
033402 (2000)

\bibitem{chj21}  N. B. Delone, and V. P. Krainov, J Opt. Soc. Am. B {\bf 8},
1207 (1991)

\bibitem{rbp421u}  R. Abrines and I.C. Percival, Proc. Phys. Soc. {\bf 88},
861 (1966); J.G. Leopold and I.C. Percival, J. Phys. B {\bf 12}, 709 (1979)

\bibitem{bdh8}  J.S. Cohen, Phys. Rev. A {\bf 26}, 3008 (1982)

\bibitem{rbp421}  C.H. Keitel and P.L. Knight , Phys. Rev. A {\bf 51}, 1420
(1995); G. Bandarage, et al., ibid. {\bf 46}, 380 (1992); M. Gajda, et al.
ibid. {\bf 46}, 1638 (1992); G.A. Kyrala, J. Opt. Soc. Am. B {\bf 4}, 731
(1992)

\bibitem{bdh}  J. Liu, S. G. Chen and D. H. Bao, Comm. Theor. Phys. {\bf 25}%
, 129 (1996)

\bibitem{chj22}  L.D. Landau, E. M. Lifishitz, {\it Quantum Mechanics. }%
(Rergamon, Oxford, 1977)

\bibitem{cfi11}  A.M. Perelomov, V.S. Popov and V.M. Teren'ev, Zh. Eksp.
Teor. Fiz. 50, 1393 (1966); M.V. Ammosov, N.B. Delone, and V.P. Krainov,
ibid 91, 2008 (1986)

\bibitem{cfiown}  T. Brakec, M.Yu. Ivanov and P. Corkum, Phys. Rev. A {\bf %
54 }, R2551 (1996)

\bibitem{newnew}  J. Chen, J. Liu, L.B. Fu and W.M. Zheng, Phys. Rev. A, to
be published as a Rapid Communication.
\end{references}
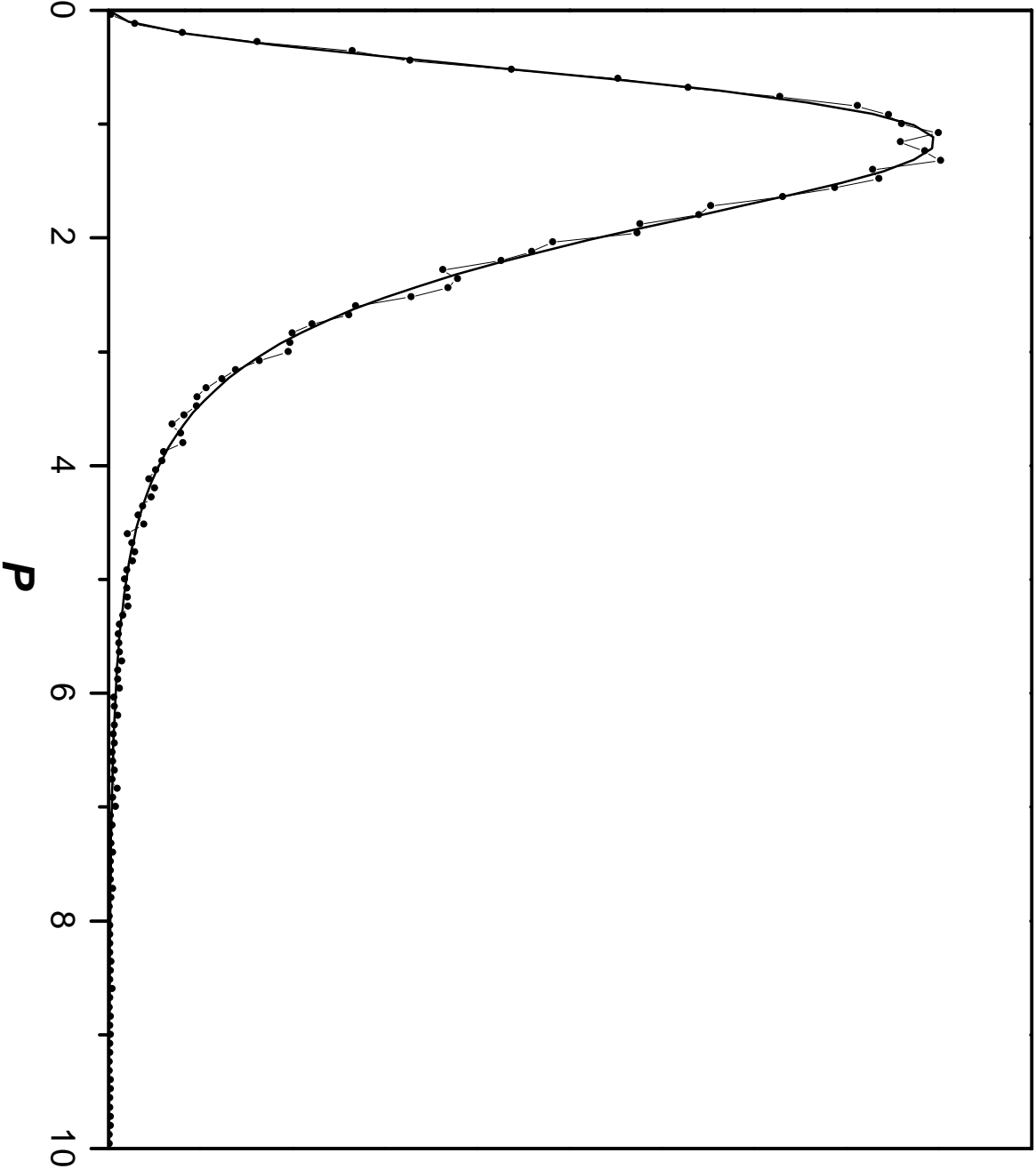
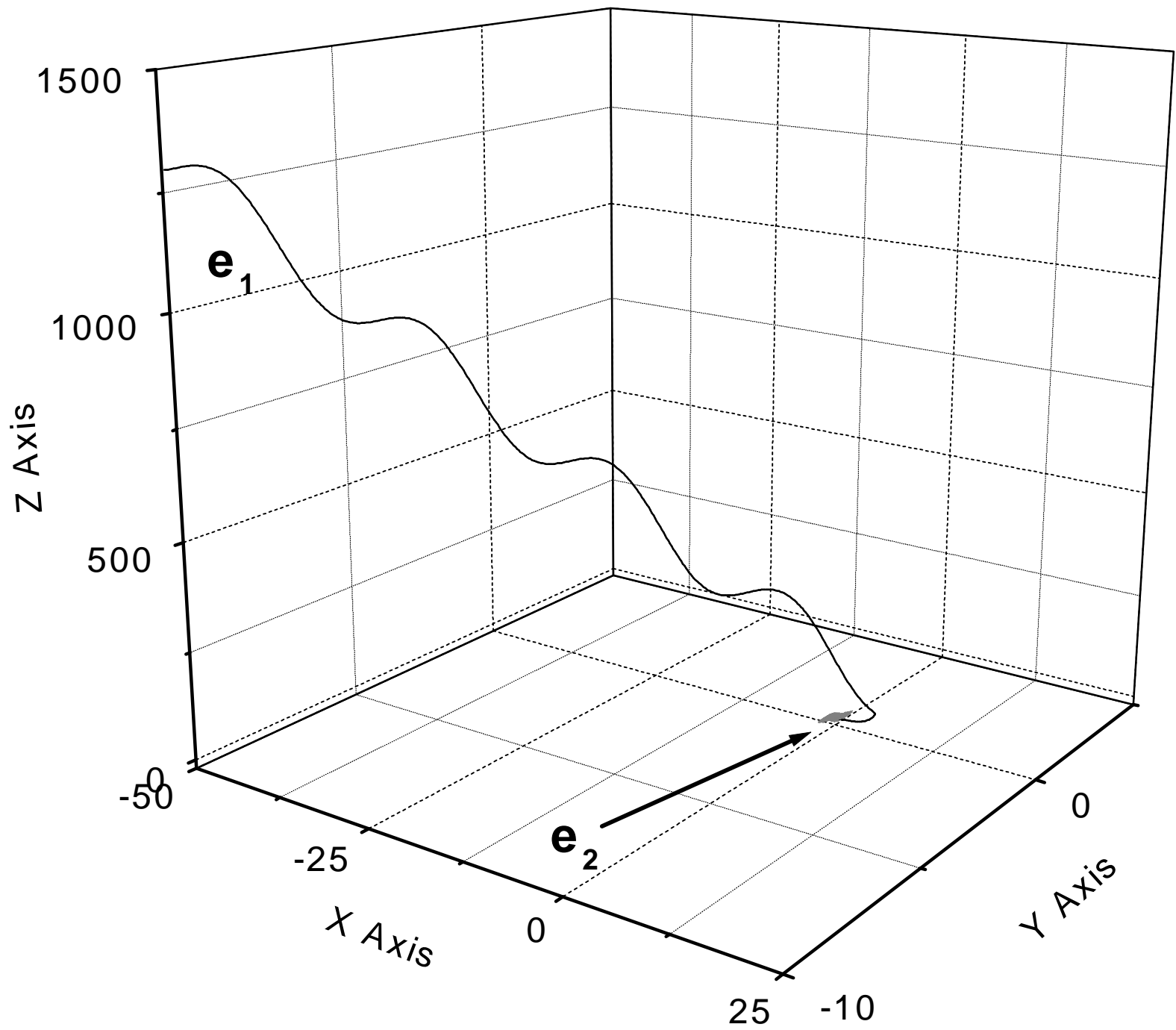
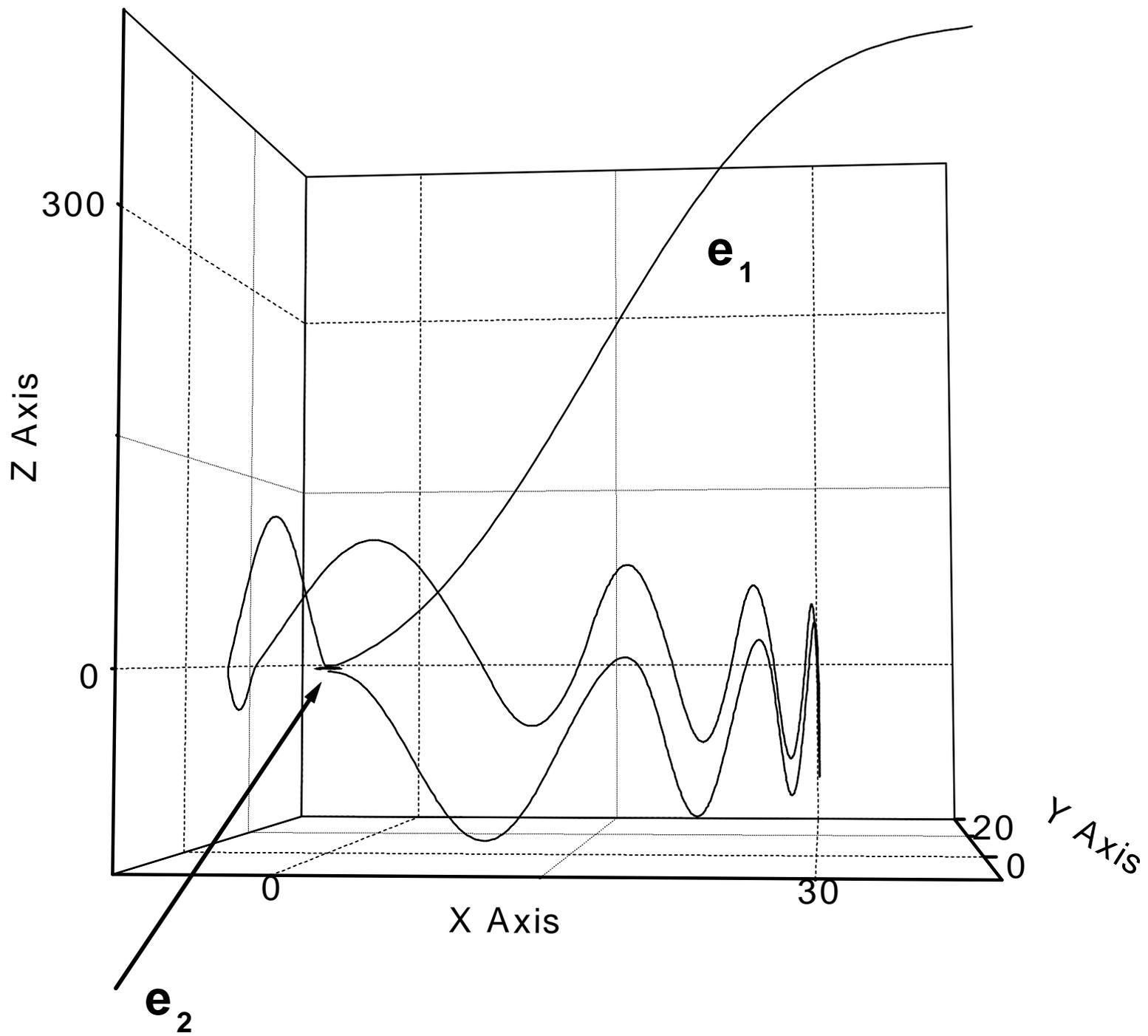
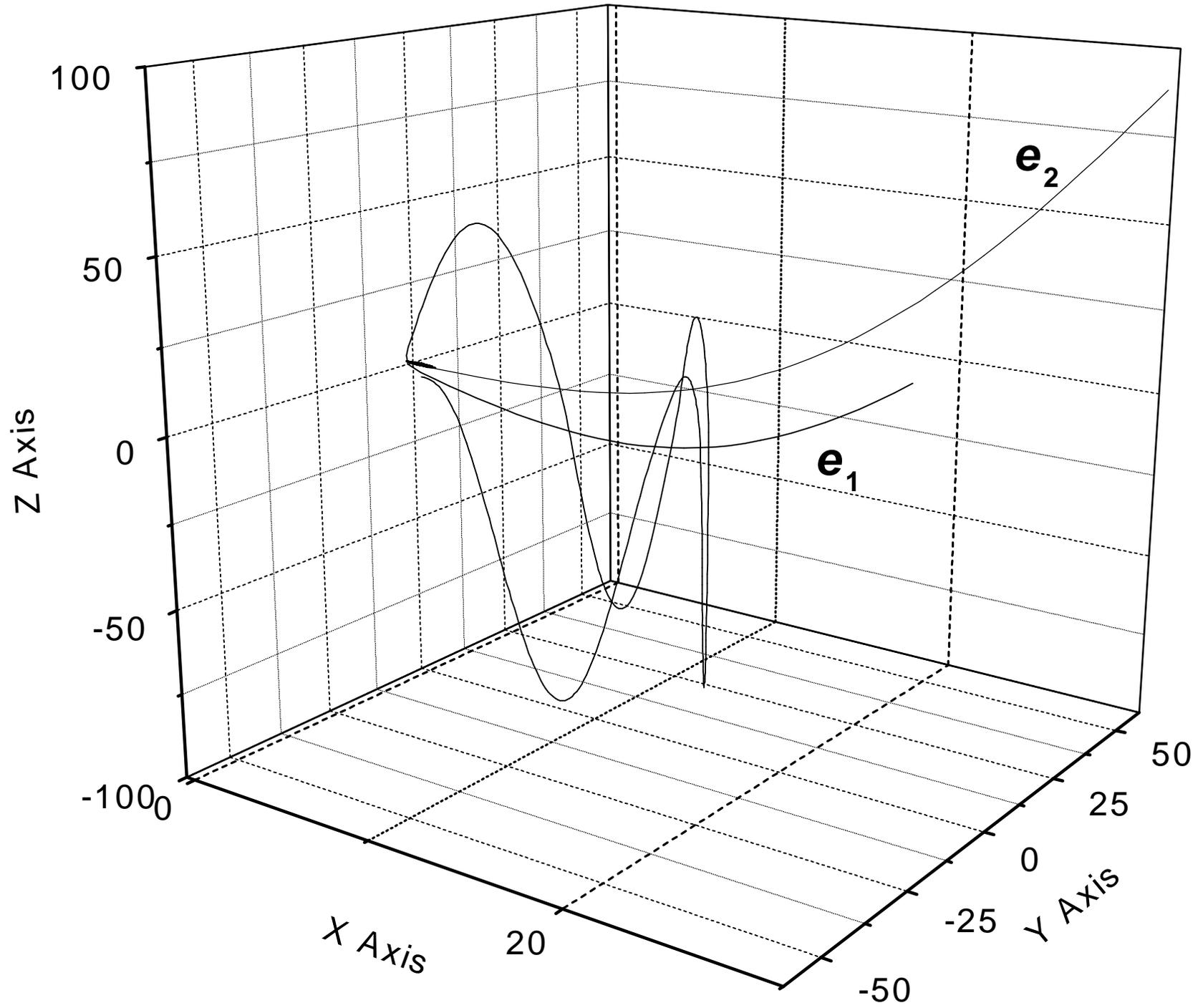
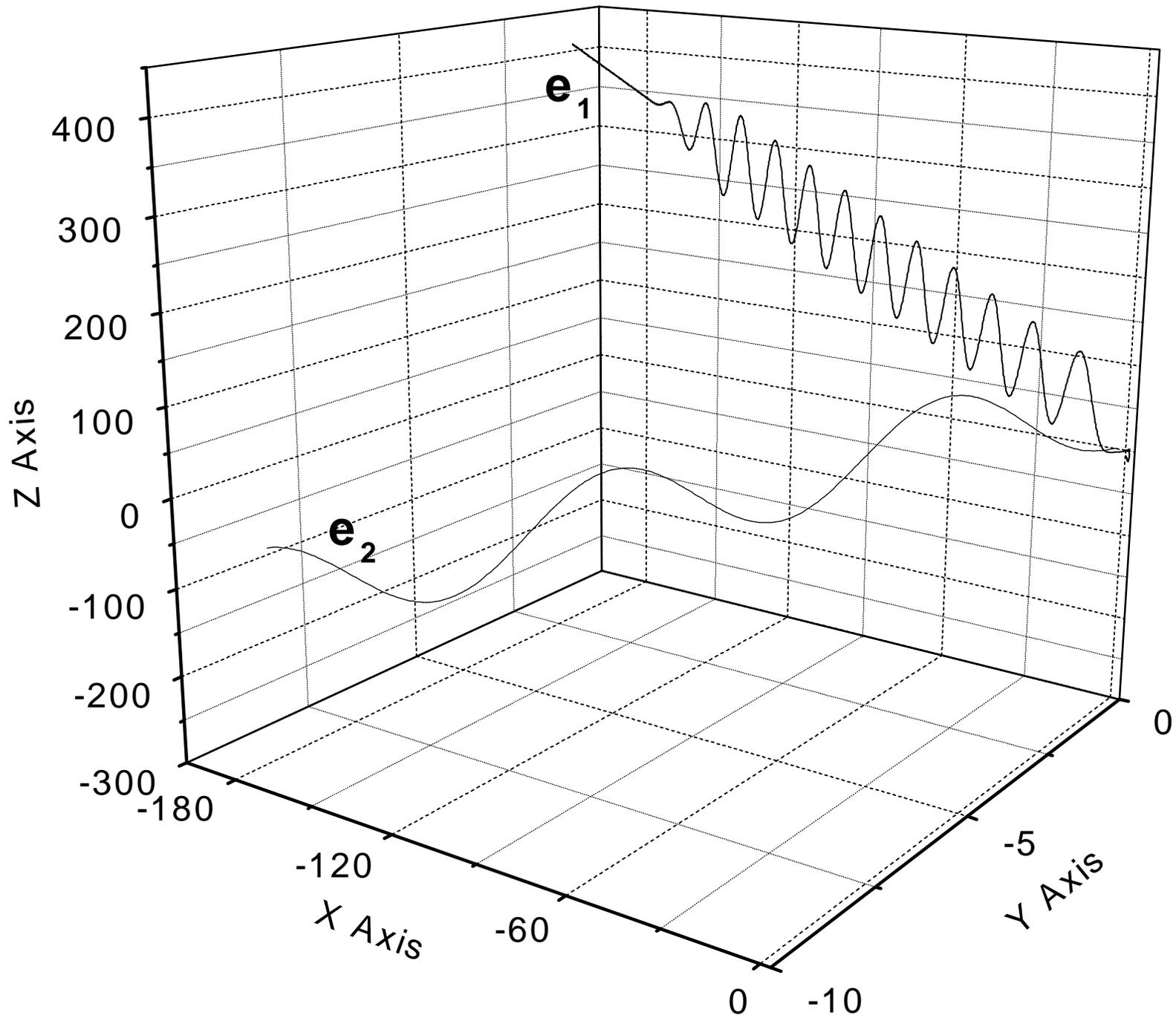
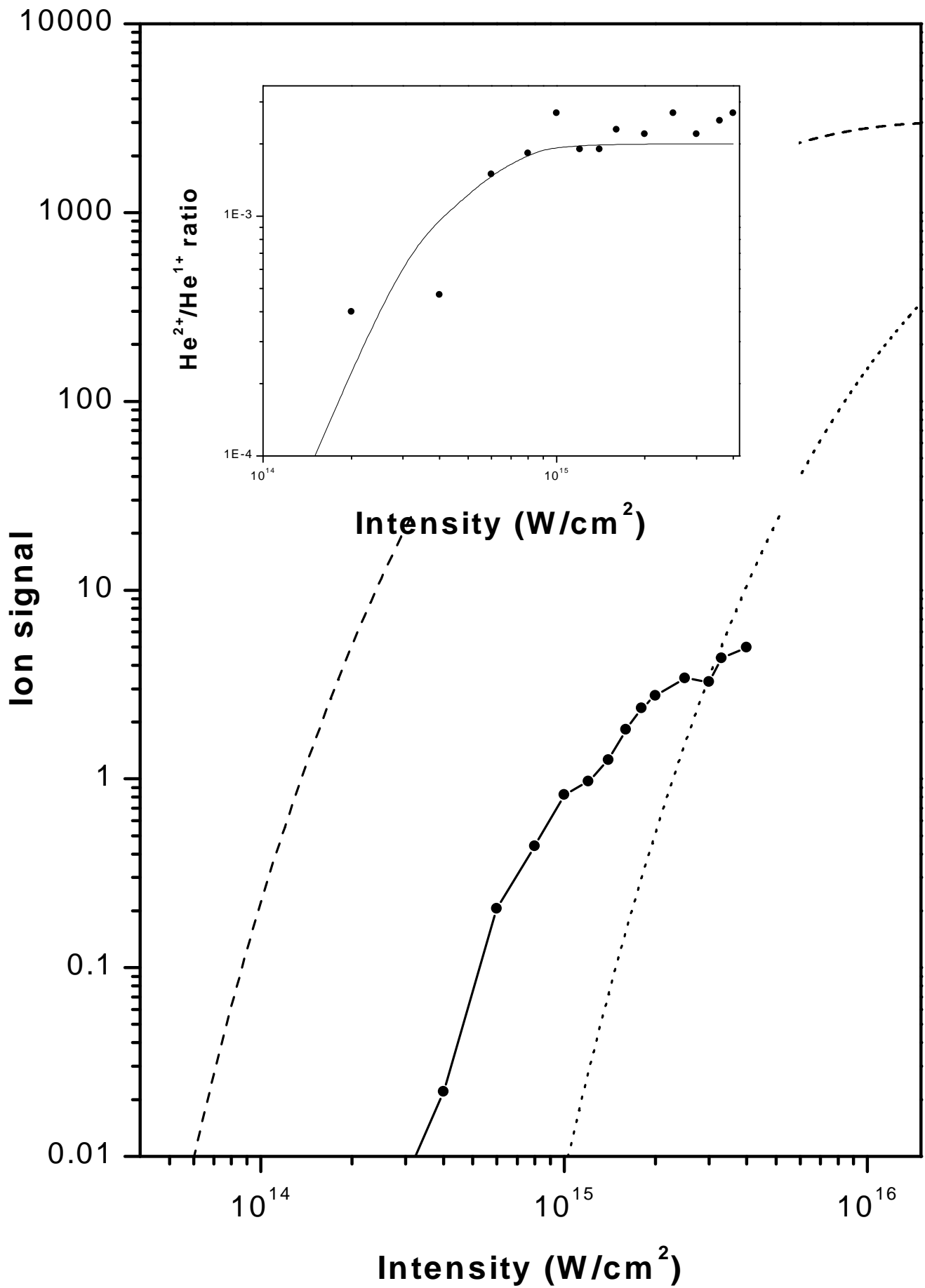
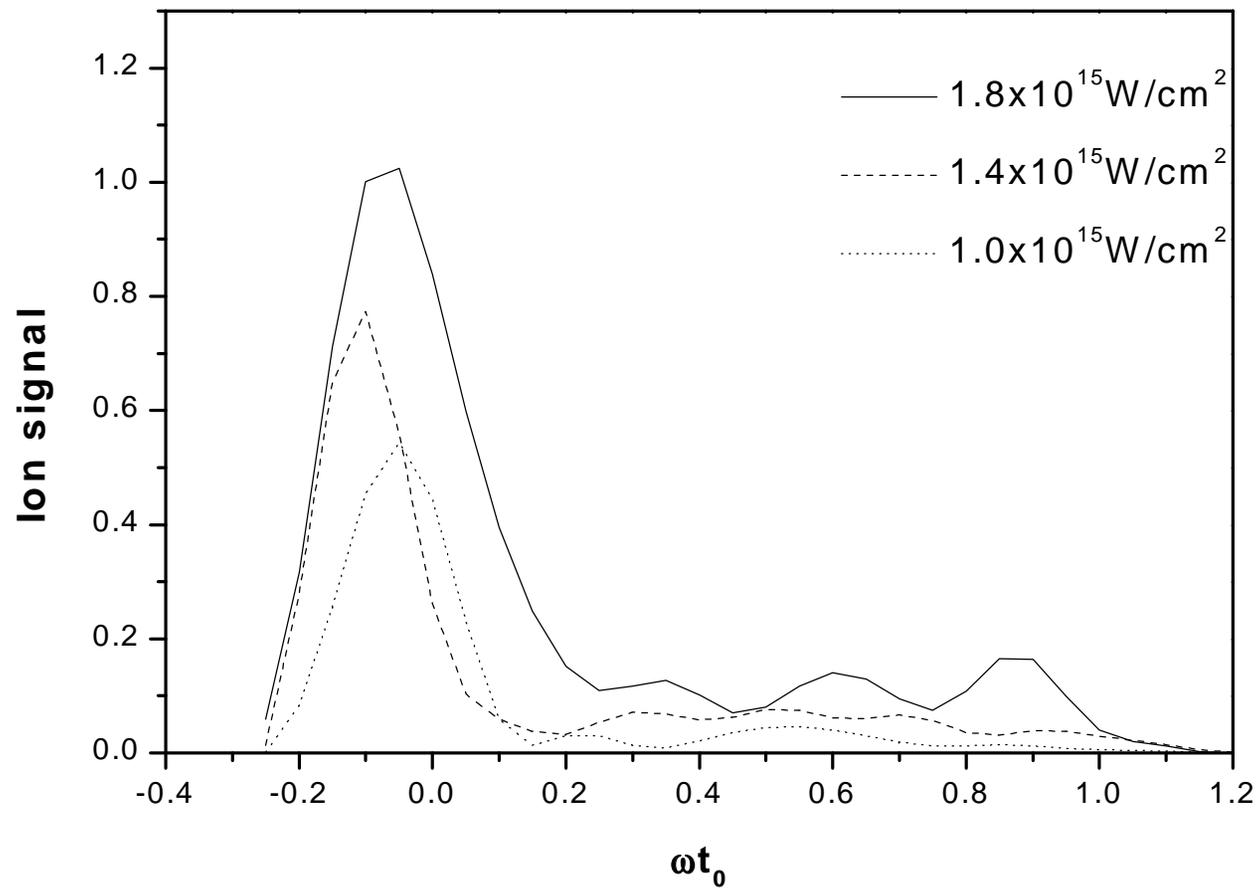
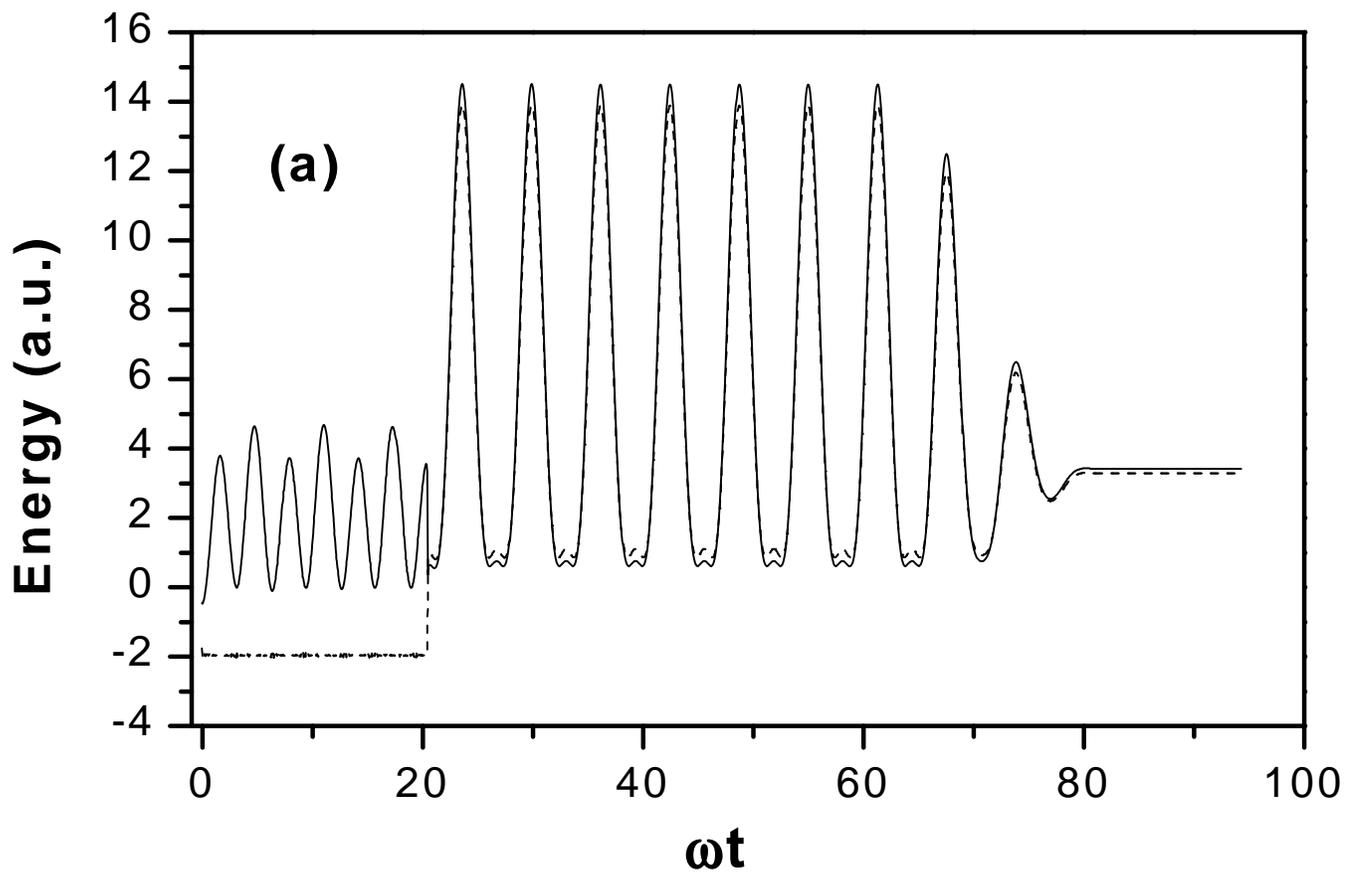
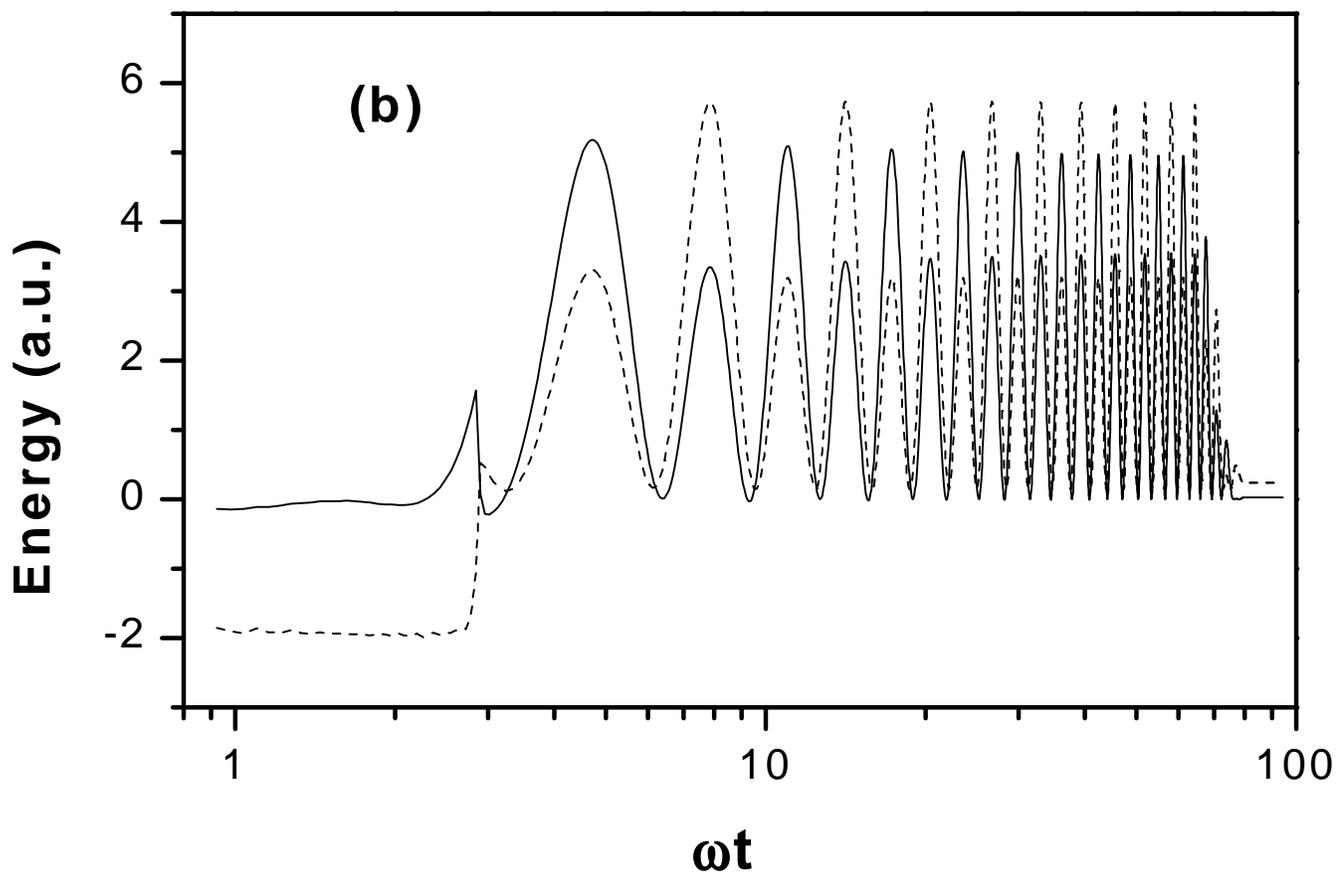
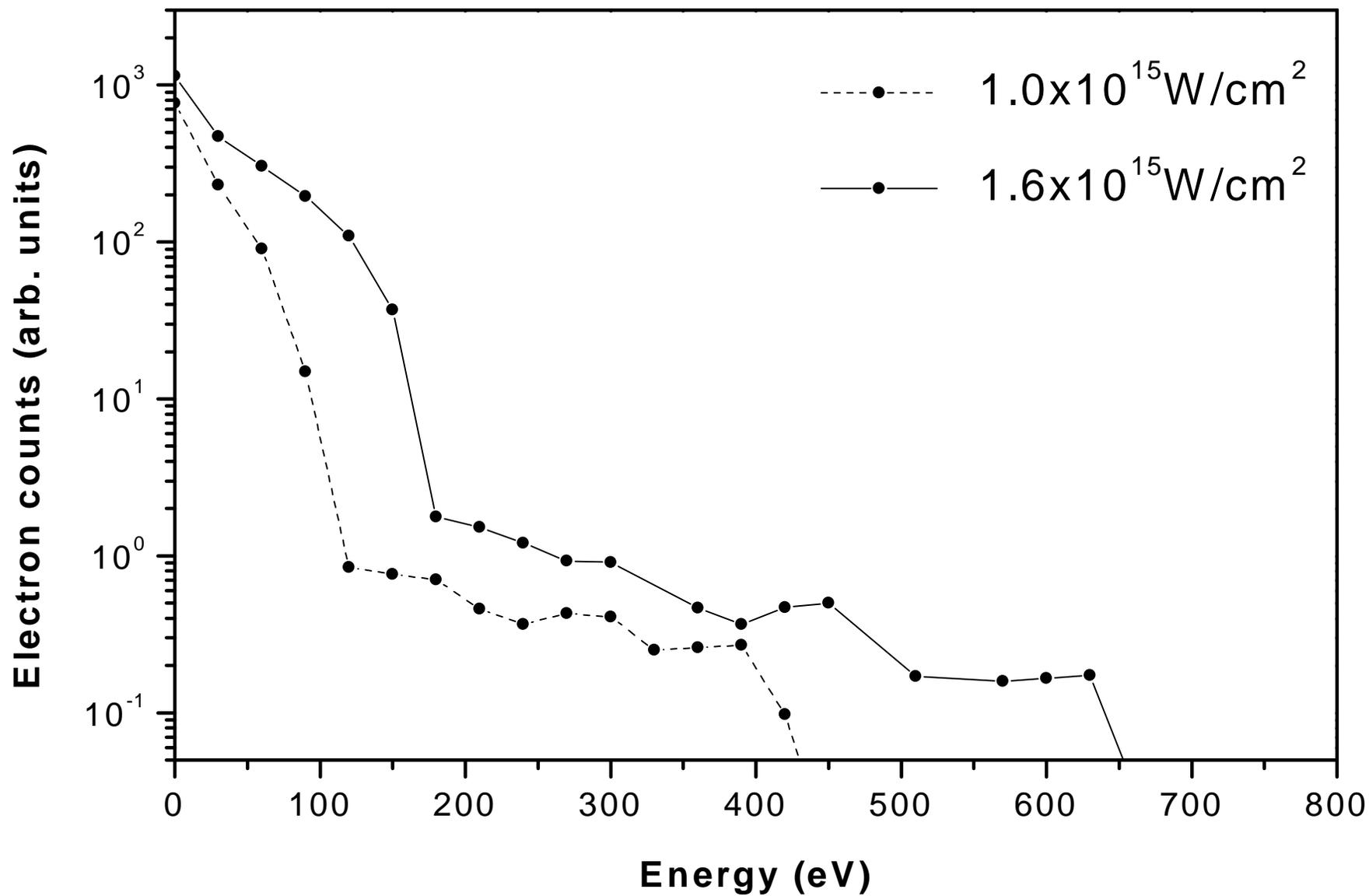
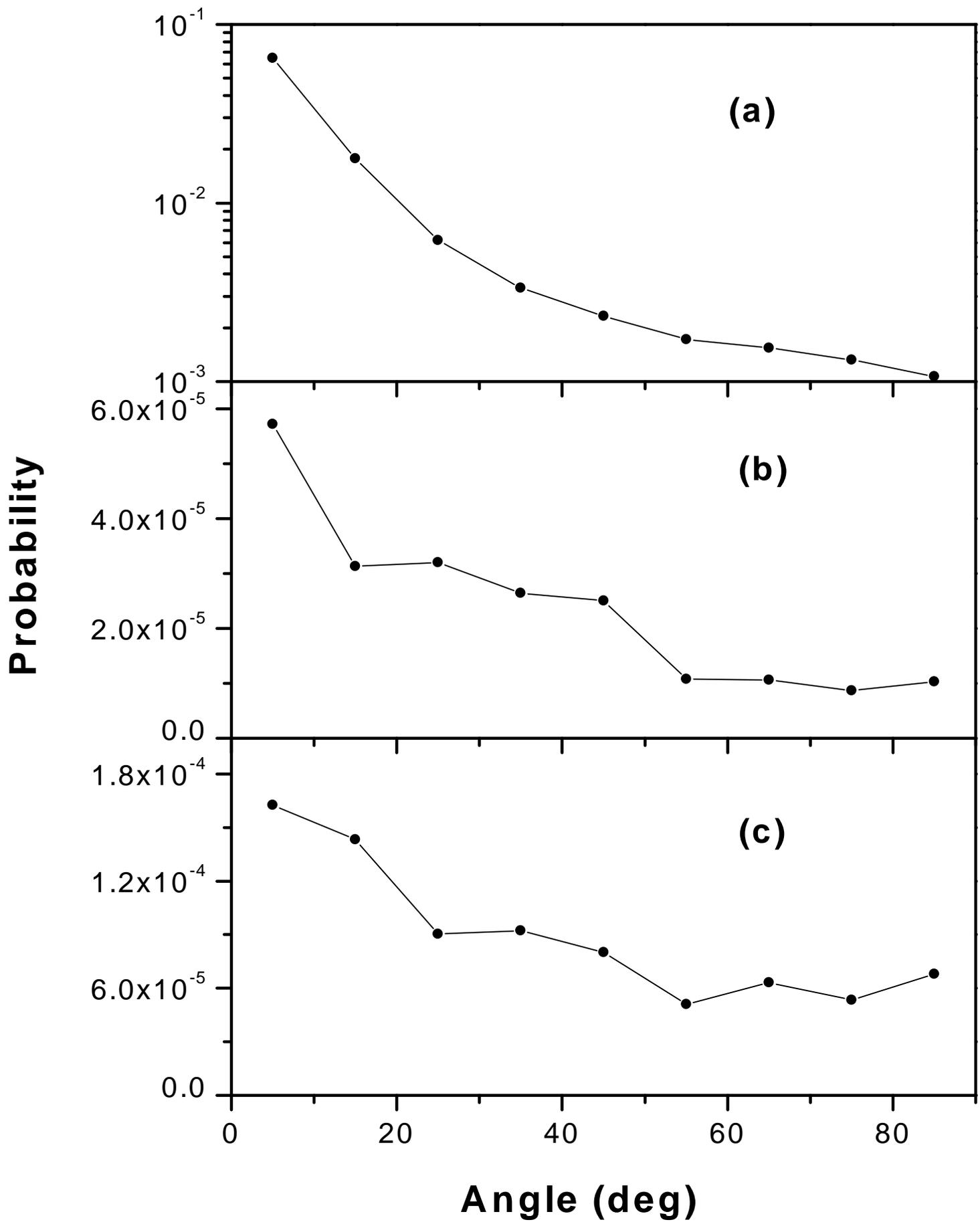
\end{document}